\documentclass[twocolumn,prl,aps]{revtex4}
\usepackage{amsmath}
\usepackage{amssymb}
\usepackage{amsfonts}
\usepackage{graphicx}

\begin{document}
\title
{Novel relativistic plasma excitations in a gated two-dimensional electron system}

\author{V.~M.~Muravev, P.~A.~Gusikhin, I.~V.~Andreev,
I.~V.~Kukushkin}
\affiliation{Institute of Solid State Physics, RAS, Chernogolovka, 142432 Russia.}

\date{\today}

\begin{abstract}
The microwave response of a two-dimensional electron system (2DES) covered by a conducting top gate is investigated in the relativistic regime for which the 2D conductivity $\sigma_{2 \rm{D}} > c/2\pi$. Weakly damped plasma waves are excited in the gated region of the 2DES.
The frequency and amplitude of the resulting plasma excitations show a very unusual dependence on the magnetic field, conductivity, gate geometry and separation from the 2DES.  
We show that such relativistic plasmons survive for temperatures up to $300$~K, allowing for new room-temperature microwave and terahertz applications.


\end{abstract}

\maketitle

Light-matter coupling between a two-dimensional electron system (2DES) and a photon microcavity has led to new studies being conducted in many research fields, such as polariton physics~\cite{Kukushkin:03, Kukushkin:06} and the behaviour of matter in the ultrastrong coupling regime~\cite{Muravev:11, Scalari:12}. Most previous studies have focused on the electrodynamic effects under the condition in which the dispersion curves for 2D plasmons and light exhibit an anticrossing. In this case a strong reduction in the resonant plasma frequency is observed due to the hybridization of plasma and light modes~\cite{Kukushkin:03, Kukushkin:06}. In a perpendicular magnetic field $B$, hybrid cyclotron-plasmon modes appear with a very unusual zigzag dependence of the frequency on the magnetic field. In some cases, cyclotron-plasmon modes and resonator photonic modes exhibit ultrastrong coupling~\cite{Gunter:09, Muravev:11, Scalari:12}. Such a regime is especially promising for observation of nonadiabatic cavity quantum electrodynamics effects~\cite{Ciuti:05}.

However, there is a second regime in which electrodynamic effects govern 2DES plasma dynamics. The conductivity $\sigma$ defines the character and rate of the Maxwellian relaxation of charge fluctuations. In bulk conductors, the excess charge density $\rho (r,t)$ relaxes without a change in the initial distribution $\rho (r,0)$ and with a decrement of $4 \pi \sigma_{3 \rm{D}}$ (Gaussian units are used in the present work unless otherwise stated): 
\begin{equation}
\rho (r,t) = \rho (r,0) \exp (-4 \pi \sigma_{3 \rm{D}} t).
\end{equation}
The two-dimensional conductivity, $\sigma_{\rm {2D}}$, of a film has units of velocity. According to electrostatic approach, when speed of light is assumed infinite, charges in a two-dimensional system relax by spreading with an effective velocity $2 \pi \sigma_{2 \rm{D}}$~\cite{Govorov:87, Dyakonov:87}. Of course, relativistic effects, such as the finite speed of light, do not enter into quasistatic results. When the film square resistance becomes smaller than $188$~$\Omega$ the velocity $2 \pi \sigma_{2 \rm{D}}$ exceeds speed of light $c$. For systems with $2 \pi \sigma_{\rm 2D} > c$, the electrostatic approach ceases to remain valid and electrodynamic effects must be taken into account. Heretofore, the dynamic response of a 2DES with such a large conductivity has not received a due attention, despite a number of interesting physical predictions~\cite{Govorov:89, Falko:89, Mikhailov:96, Mikhailov:04}, including the strong enhancement of radiative decay of collective excitations~\cite{Mikhailov:96, Mikhailov:04} and the emergence of a new family of weakly damped relativistic plasma excitations~\cite{Falko:89}. 

In this letter, we experimentally address the electrodynamic response of a gated 2DES in the regime of $\sigma_{2 \rm{D}} > c/2\pi$. A new set of weakly damped relativistic plasma modes arises in this regime. These modes are excited even at frequencies $\omega < 1/\tau$ ($\tau$ is the relaxation time for charge carriers), where ordinary 2D plasmons are overdamped~\cite{Allen:77, Heitmann:91, Andreev:14}. We show that the plasmon magneto-dispersion exhibits a suppressed magnetic field dependence, indicating the strong polaritonic nature of such excitations. The excitation, localization, and spectrum of the modes are connected to a conducting gate that covers the 2DES. Finally, we demonstrate that the plasma mode survives up to $300$~K, allowing for new tunable room-temperature microwave and subterahertz applications~\cite{Knap, Popov:11, Muravev:12, Aizin:13}.

\begin{figure}[!t]
\includegraphics[width=0.47 \textwidth]{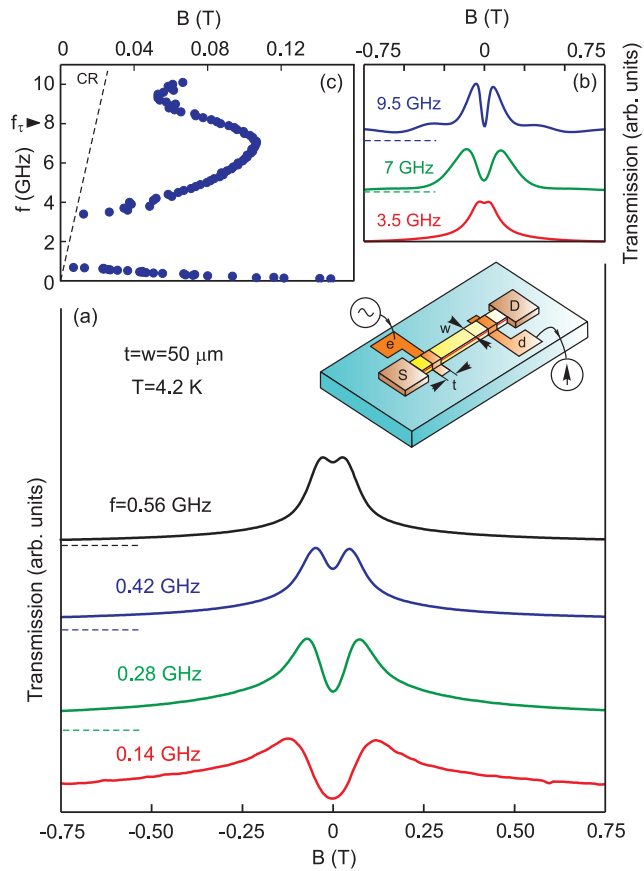}
\caption{(a-b) Magnetic field dependencies of microwave transmission of a sample with $t=W=50$~$\mu$m and an electron density of $1{.}4 \times 10^{11}/{\rm cm}^2$ ($2\pi\sigma_{2\rm{D}}/c = 8$) measured for several microwave frequencies. Curves are offset for clarity. The dashed lines indicate zero signal level in the absence of input power. (c) Two branches of the magnetic-field position of the absorption resonances versus excitation frequency. The dashed line represents the cyclotron resonance. The inset shows a schematic view of the sample geometry.}
\label{1}
\end{figure} 

We present experimental results on a 2DES fabricated from modulation-doped GaAs/AlGaAs heterostructures. The density in different samples ranged from $n_s=1{.}4 \times 10^{11}$ to $44 \times 10^{11}/{\rm cm}^2$. The electron mobility at $T=4{.}2$~K varies from $10^5$ to $6 \times 10^6~{\rm cm}^{2}/{\rm Vs}$ (see Supplementary Material \textrm{I}). Stripe-shaped mesas with widths $w=50$, $100$, and $150$~$\mu$m and a length of $L=680$~$\mu$m were fabricated from these heterostructures. The stripe terminates in source and drain ohmic contact. Near these ohmic contacts at a distance of $40$~$\mu$m, identical $20$~nm Cr and $100$~nm Au gate fingers, which run across the stripe, are deposited on top of the mesa (see Fig.~1(a) for a schematic drawing). Gates with widths ($t$) of $10$, $20$, $40$, $50$, and $80$~$\mu$m were investigated. The distance between the 2DES and top conducting gate layer $h$ varies between $h=50$~nm and $400$~nm for different structures. Microwave radiation with frequencies from $0{.}01$ to $40$~GHz and modulated at $1$~kHz was guided into the cryostat with a coaxial cable and transferred to the excitation gate $e$ via an impedance-matched coplanar waveguide transmission line. The ground planes of the coplanar waveguide are connected to adjacent source and drain contacts of the 2DES stripe. The microwave power at the entrance to the coax cable does not exceed $10$~$\mu$W. The gate $e$ serves to excite plasma excitations in the 2DES, while gate $d$ is used to detect the induced resonant AC potential. The output microwave power from gate $d$ is detected by a Schottky diode with a preamplifier using a standard lock-in technique. The sample is placed in a helium cryostat at the centre of a superconducting coil. Experiments are carried out by applying a magnetic field ($0$~-- $4$~T) perpendicular to the sample surface at temperatures $T=4{.}2$~-- $300$~K.  

Figure~1(a-b) shows the magnetic field dependence of the microwave transmission detected at gate $d$, while the excitation power at gate $e$ is kept fixed. In these experiments gates $e$ and $d$ have identical geometry with $t=W=50$~$\mu$m, and separation between the top gate and the 2DES is $h=200$~nm. The traces have been offset vertically for clarity and the dashed lines indicate the zero signal level in the absence of the input microwave power. Each curve shows a resonance that is symmetrical with respect to zero magnetic field. The resonances were recorded for different microwave frequencies $f$ in the range $0{.}05 - 10$~GHz (Fig.~1(a-b)). The resulting magnetic field behaviour of the resonance is displayed in Fig.~1(c). The magnetodispersion has two branches. 
The magnetic field and electron density behaviour of the resonance unambiguously indicates its plasmonic nature (see Supplementary Material \textrm{IV}). However, a number of features of this resonance are drastically different from those of all known plasma modes observed thus far. (i) The mode is observed in the frequency range $f < f_{\tau}=1/\pi \tau$ ($\tau$ corresponds to the transport relaxation time for the sample under study), where ordinary plasma excitations are overdamped~\cite{Allen:77, Heitmann:91, Andreev:14}. (ii) The upper magneto-dispersion curve lies under the cyclotron resonance line, and exhibits a zigzag shaped behaviour at higher frequencies. Such behavior is typical for plasmon polariton excitations~\cite{Kukushkin:03}. (iii) Frequencies of all known plasma excitations, calculated for the investigated structure geometry and parameters are situated over $20$~GHz at $B=0$~T (see Supplementary Material \textrm{III}). 
All these facts suggest that we have observed a novel polaritonic plasma excitation. We assert that the observed plasmon mode is excited and localized in the gated region of the 2DES. Each of the gates $e$ and $d$ defines independent plasmonic resonators for the mode under study.
We experimentally verified that the presence of contacts and the distance between gates $e$ and $d$ have no significant effect on the properties of the observed plasmons. This point of view will be corroborated in the remainder of the manuscript by investigating the mode behaviour as a function of temperature, density and geometry in further detail. 

\begin{figure}[!t]
\includegraphics[width=0.47 \textwidth]{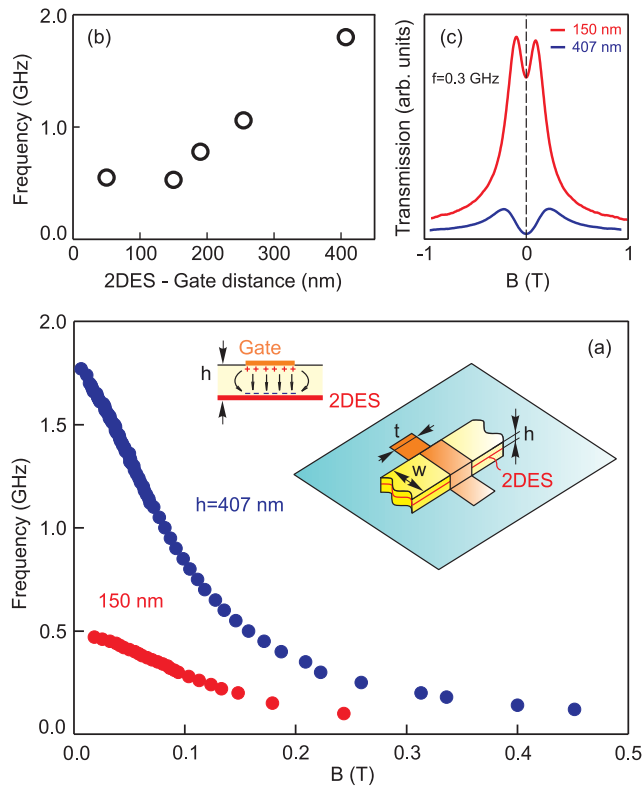}
\caption{(a) Magnetic field dependencies of the plasma mode of two microstructures with different Gate~- 2DES separation. Structures have the same values for $t$ and $w$, which are equal to $40$~$\mu$m and $50$~$\mu$m, respectively. Schematic geometry with designated dimensions is shown in the inset.
(b) Zero-field limit frequency of the plasma mode as a function of Gate~- 2DES separation.
(c) Magnetic field dependencies of microwave transmission of the samples with different Gate separations at a frequency of 0.7~GHz. The electron density in all structures is in the range $n_{s}=(1.7 - 2.1) \times 10^{11}$~cm$^{-2}$.}
\label{4}
\end{figure}  

\begin{figure}[!t]
\includegraphics[width=0.47 \textwidth]{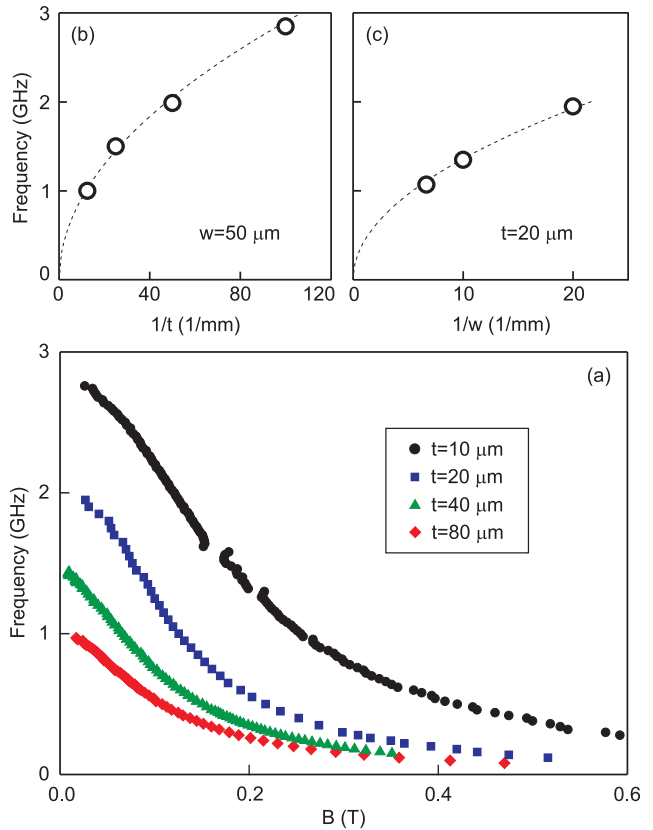}
\caption{(a) Magnetic field dependence of the plasma mode in a resonator with width $w=50$~$\mu$m and different lengths $t$. The electron density in the structure is $n_{s}=4.3\times 10^{11}$~cm$^{-2}$. The top figures plot the frequency of the plasma mode as a function of the resonator length $t$ (b) as well as the resonator width $w$ (c).   
} 
\label{5}
\end{figure} 

The conducting top gate over the 2DES plays a crucial role in the plasma mode formation and excitation. All further experimental results will be presented for the low-frequency plasma branch. Quantitatively, all results stay the same for the high-frequency branch, and will be published elsewhere. In Fig.~2(a) the magnetic field dependencies of the plasmon resonance frequencies are plotted for two structures with different distances $h$ between the gate and 2DES (inset to Fig.~2(a)). Both structures have identical electron densities of $n_s=1.9\times 10^{11}$~cm$^{-2}$ and similar values for the 2D conductivity $2 \pi \sigma_{2\rm{D}}/c = 3$. The distance $h$ has a strong effect on the plasmon magneto-dispersion. The magnetic field dependence of the plasma excitation frequency can be phenomenologically described by~\cite{Volkov:88, Aleiner:95}:
\begin{equation}
\omega^2=\omega_0^2/(1 + \omega_c^2/\omega_B^2),
\label{CR_L}
\end{equation}
where $\omega_0$ denotes the mode frequency at $B=0$~T, while $\omega_B$ characterizes the magnetodispersion slope. To better understand the properties of the plasma waves involved in our investigation, the experiment shown in Fig.~2(a) was repeated for a set of structures with $h=50$, $150$, $190$, $255$, and $407$~nm. The mode frequencies at zero magnetic field obtained using the approximation of Eq.~(\ref{CR_L}) are plotted in Fig.~2(b) as a function of the distance between the gate and 2DES. We assume that the observed mode has an axisymmetric dipolar nature with strong electric field concentrated in the Gate~- 2DES slot space~\cite{Chaplik:72, Fetter:86}. The mode formation may qualitatively be understood as follows. Plasma charge fluctuations in the 2DES are screened by the accumulation of charges with an opposite sign in the conducting gate (Fig~2(a)). The separated charges form oscillating dipoles that radiate electromagnetic waves into free space. Such a wave appears to be a photon cloud of the hybrid plasmon polariton excitation under study. To further support this point of view we studied how Gate~- 2DES separation influences the resonance amplitude. Fig.~2(c) shows that as the separation increases from $h=150$~nm to $407$~nm a suppression of more than one order of magnitude is observed in transmission. Moreover, additional experiments prove that the resonance completely disappears whenever the top gate is removed or the 2DES under the gate is depleted. This confirms the fact that the conducting gate serves to excite the plasmon excitation and to form a plasmonic resonator, where most of the mode electromagnetic energy is stored.

In order to substantiate our vision of the plasmon mode location and dispersion, we conducted a series of experiments with samples of different geometry. We found that only the dimensions of the conducting gate, $t$ and $w$, determine the mode frequency. Figure~3(a) shows a plot of the magneto-dispersions obtained for the plasmonic resonator with $w=50$~$\mu$m and different values of $t$ ranging from $10$~$\mu$m to $80$~$\mu$m. The structure under study had an electron density of $n_{s}=4.3\times 10^{11}$~cm$^{-2}$ with $2\pi\sigma_{2\rm{D}}/c = 5.5$, and a Gate~- 2DES distance of $h=200$~nm. In Fig.~3(b), we extracted the zero-field mode frequency using Eq.~(\ref{CR_L}) and plotted the dispersion curves. The dashed line represents a square root fit, which is in good agreement with the experimental data. Samples with different 2DES stripe widths $w$ ($t=20$~$\mu$m) were also investigated. It is shown in Fig.~3(c) that the mode frequency in the zero field limit has the same square root dependence on the value of $1/w$. As a result, we obtain the following relation for the mode dispersion $f_0 \sim 1/\sqrt{w\cdot t}$. For a square gate geometry, $w=t$, and the plasmon excitation has a linear dispersion analogous to the spectrum of an acoustic plasmon in bilayer systems~\cite{Muravev:07}.

\begin{figure}[!t]
\includegraphics[width=0.47 \textwidth]{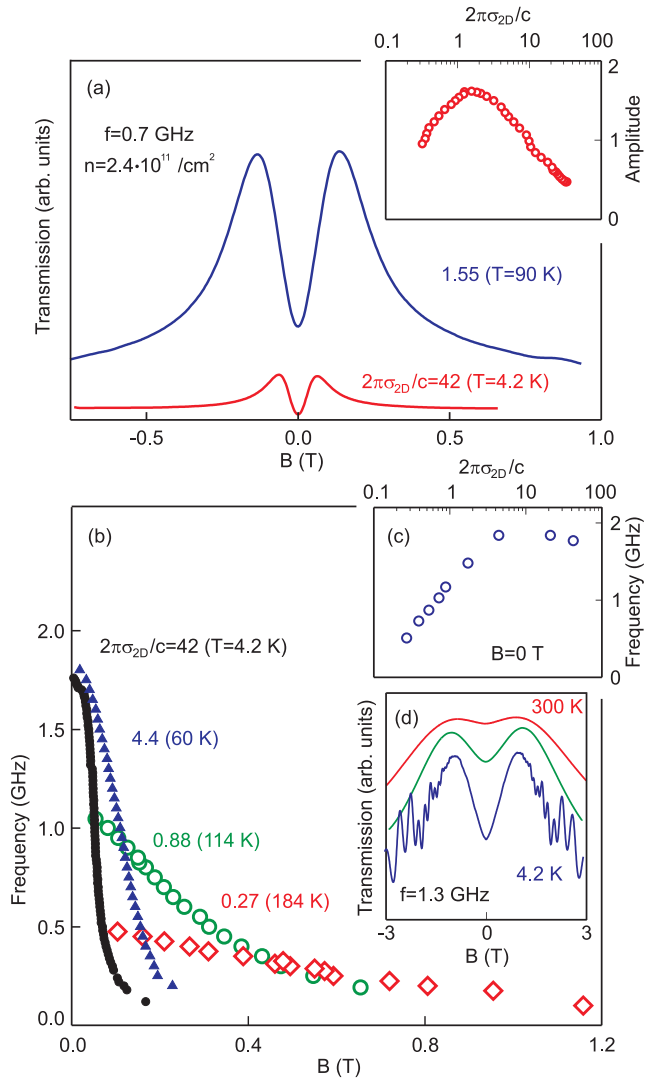}
\caption{(a) Magnetic field dependence of the microwave transmission of a sample with $t=W=50$~$\mu$m, mobility $\mu=6\times 10^6$~cm$^2$/V$\cdot$s, and electron density of $2{.}4 \times 10^{11}/{\rm cm}^2$ measured for two temperatures $T=4.2$~K and $T=90$~K at a frequency of $0.7$~GHz. The inset shows a plot of the resonance amplitude as a function of the electrodynamic parameter $2\pi\sigma_{2\rm{D}}/c$. (b) Magnetodispersion of the plasma mode for the same structure at different temperatures. (c) Zero-field extrapolation of the plasma mode frequency as a function of $2\pi\sigma_{2\rm{D}}/c$. (d) Magnetic field dependencies of the microwave transmission of a sample with a record electron density of $44 \times 10^{11}/{\rm cm}^2$ measured for three temperatures $T=4.2$~K, $160$~K, and $300$~K at a frequency of $1.3$~GHz.
}
\label{2}
\end{figure} 

Figure~4(a) shows the magnetic field dependencies of the transmission for a sample with a fixed geometry at different temperatures. The structure has a mobility of $6\times 10^6$~cm$^2$/V$\cdot$s and electron density $n_s=2.4\times 10^{11}$~cm$^{-2}$ ($T=4{.}2$~K). The separation between the top gate and the 2DES is $h=400$~nm. The conductivity $\sigma_{2 \rm{D}}$ and density of the structure were measured by a four-probe transport technique on a nearby situated Hall bar. These auxiliary experiments show that the 2DES density is largely unaffected by temperature up to $T=200$~K. Traces of Fig.~4(a) demonstrate that $\sigma_{2\rm{D}}$ strongly influences the amplitude and magnetic field position of the plasmon resonance. The inset of Fig.~4(a) shows the conductivity dependence of the resonance amplitude for $f=0{.}7$~GHz. The resonance reaches its maximum around the value of $2\pi\sigma_{2\rm{D}}/c\sim 1$, and abruptly disappears at smaller conductivities~\cite{Mikhailov:04}. Figure~4(b) plots the plasmon magnetodispersions for the set of temperature values from $T=4{.}2$~K up to $188$~K. Corresponding values of $2\pi\sigma_{2\rm{D}}/c$ measured by a transport technique are displayed near the data points. The frequency $\omega_B$ from Eq.~(\ref{CR_L}) raises as the temperature is increased up to $90$~K ($2\pi\sigma_{2\rm{D}}/c > 1$), whereas the mode frequency $\omega_0$ at zero magnetic field remains approximately constant. This behaviour contradicts the well-established fact for ordinary 2D and 1D plasmons that the excitation frequency does not depend on the 2DES temperature~\cite{Kukushkin:06, Andreev:14}. In the limit $2\pi\sigma_{2\rm{D}}/c < 1$, the plasmon mode amplitude and frequency $\omega_0$ start to drop rapidly (insets of Fig.~4(a-b)), and near $2\pi\sigma_{2\rm{D}}/c = 0{.}3$ the mode disappears. Taking into account the fact that in the electrostatic approach, excess charge in a 2DES spreads with an effective velocity $2 \pi \sigma_{\rm 2D}$. The present experiments imply that the observed plasmon mode has an essentially electrodynamic nature directly related to relativistic effects such as the finite speed of light. A relativistic plasmon excitation occurs when $\sigma_{2 \rm{D}} > c/2\pi$.

To further prove this point of view, we conducted experiments in the same geometry on a structure with record electron density of $n_s=44\times 10^{11}$~cm$^{-2}$. The structure has $2\pi\sigma_{2\rm{D}}/c = 13.2$ ($\omega \tau = 0.15$) at $T=4{.}2$~K, and $2\pi\sigma_{2\rm{D}}/c = 2.6$ ($\omega \tau = 0.017$) at $T=300$~K. In Fig.~4(d), the magnetic field dependence of the transmission between two identical plasmonic resonators is plotted for $T=4{.}2$~K, $160$~K, and $300$~K. One can see that although the plasmon resonance significantly broadens with an increase in temperature, it is still present at room temperature $T=300$~K despite the fact that $\omega \tau \ll 1$. 


In conclusion, we have investigated the microwave absorption of a gated two-dimensional electron system (2DES). We have identified two resonant plasma modes at frequencies $\omega < 1/\tau$, where ordinary plasma waves are overdamped. The modes appear to be essentially relativistic in nature. This claim is corroborated by a very peculiar dependence of the resonance frequency on magnetic field, and the fact that the modes exist only when $2\pi\sigma_{2\rm{D}}/c>1$. The experimental results shown here build a thorough, qualitative picture of this phenomenon; however, a detailed theoretical treatment is highly desirable. Most importantly, we demonstrate that the relativistic plasma excitation survives in high conductivity structures at temperatures up to $300$~K, opening an avenue for the long-awaited realization of terahertz plasmon circuits based on semiconductor heterostructures. 

We thank V.~Volkov for stimulating discussions. We also gratefully acknowledge financial support from the Russian Scientific Fund (Grant No.~14-12-00693).

\end{document}


\title{Supplementary Material for\\ ``Novel relativistic plasma excitations in a gated two-dimensional electron system''}

\author{V.~M.~Muravev, P.~A.~Gusikhin, I.~V.~Andreev,
I.~V.~Kukushkin}
\affiliation{Institute of Solid State Physics, RAS, Chernogolovka, 142432 Russia.}

\date{\today}\maketitle

\section{\textrm{I}. Semiconductor structures used in experiments}
\begin{center}
\begin{tabular}{|c|c|c|c|c|}
\hline
\rule{0pt}{3ex}
Number & $h$~(nm) & $n_s$~($10^{11}$~cm$^{-2}$) & $\mu$~($10^6$~cm$^2$/V$\cdot$s) & $2\pi\sigma_{\rm 2D}/c$\\
\hline
\rule{0pt}{3ex}
1 & 190 & 1.4 & 1.9 & 8.0\\
\hline
\rule{0pt}{3ex}
2 & 407 & 1.7 & 0.5 & 2.6\\
\hline
\rule{0pt}{3ex}
3 & 255 & 1.8 & 0.8 & 4.1\\
\hline
\rule{0pt}{3ex}
4 & 190 & 1.7 & 1.3 & 6.5\\
\hline
\rule{0pt}{3ex}
5 & 150 & 2.1 & 0.5 & 3.4 \\
\hline
\rule{0pt}{3ex}
6 & 50 & 1.9 & 0.7 & 4.0 \\
\hline
\rule{0pt}{3ex}
7 & 200 & 4.3 & 0.4 & 5.5\\
\hline
\rule{0pt}{3ex}
8 & 400 & 2.4 & 6.0 & 42.0\\
\hline
\rule{0pt}{3ex}
9 & 5 QWs between 200 and 440 & 44 & 0.1 & 13.2\\
\hline
\end{tabular}
\end{center}

The table summarizes all structures used throughout our experiments. Structures $\#1$ and $\#4$ had a single AlGaAs/GaAs heterojunction, while all other semiconductor structures had a single $20$~nm quantum well (QW). For structure $\#7$ we conducted special research with QWs of widths $20$~nm, $25$~nm and $30$~nm that demonstrated insensitivity of novel plasmon mode to QW architecture.

\section{\textrm{II}. Estimate of the mode localization length}

In Fig.~1S we present the magnetic field dependence of the microwave transmission at $f=0.8$~GHz for two structures with different distances $L=0.7$~mm and $0.3$~mm between gates $e$ and $d$. In these experiments the gates $e$ and $d$ have $t=20$~$\mu$m and $W=50$~$\mu$m. The structure $\#7$ under study had an electron density of $n_{s}=4.3\times 10^{11}$~cm$^{-2}$ with $2\pi\sigma_{2\rm{D}}/c = 5.5$. Fig.~1S shows that as the distance $L$ increases a strong suppression of resonance amplitude is observed in transmission. If we assume the exponential decay $e^{-L/L_{\rm loc}}$ of the plasmon field along the 2DES stripe, we could estimate the characteristic scale of plasmon localization as $L_{\rm loc} \approx 0.8$~mm. Inset to Fig.~1S shows a plot of the magnetodispersions obtained for the distances between plasmonic resonators of $L=0.7$~mm and $0.3$~mm. As one can see, distance between the resonators does not affect the magnetodispersion. That allows us to conclude that two resonators are decoupled. 

\begin{figure}[!h]
\hfill
  \includegraphics[width=0.45\textwidth]{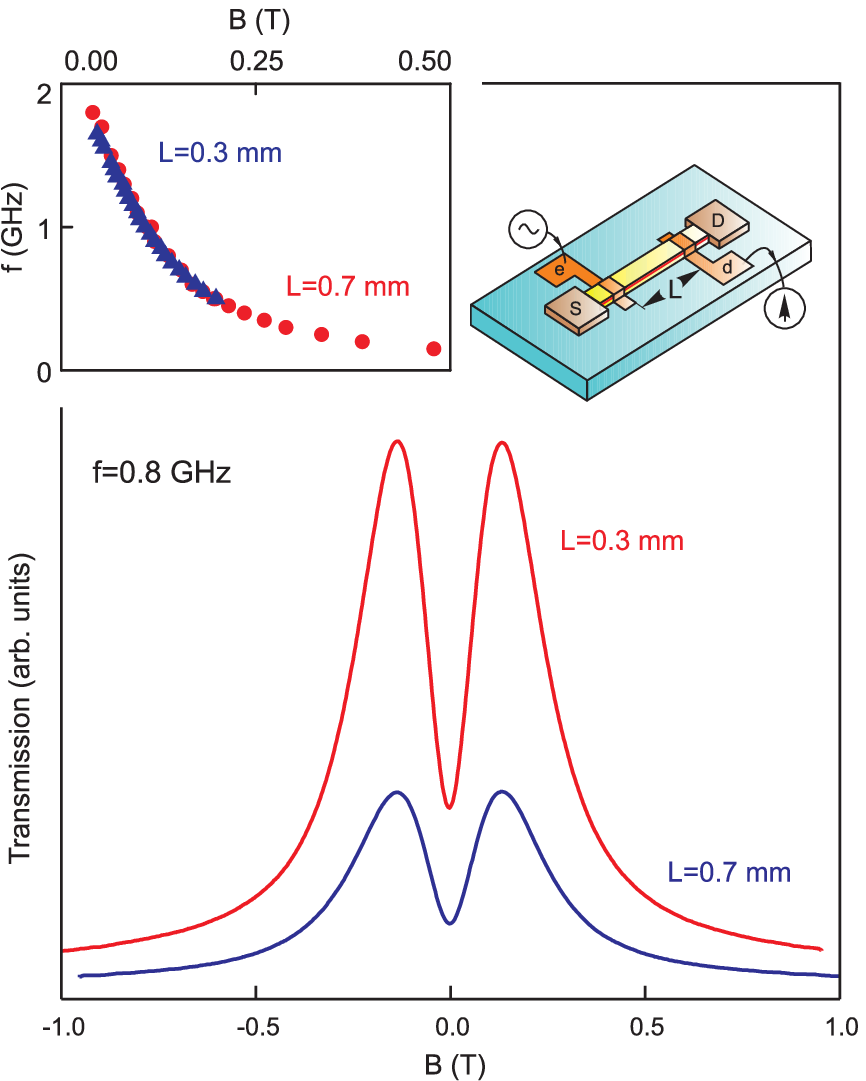}\hfill
  \includegraphics[width=0.47\textwidth]{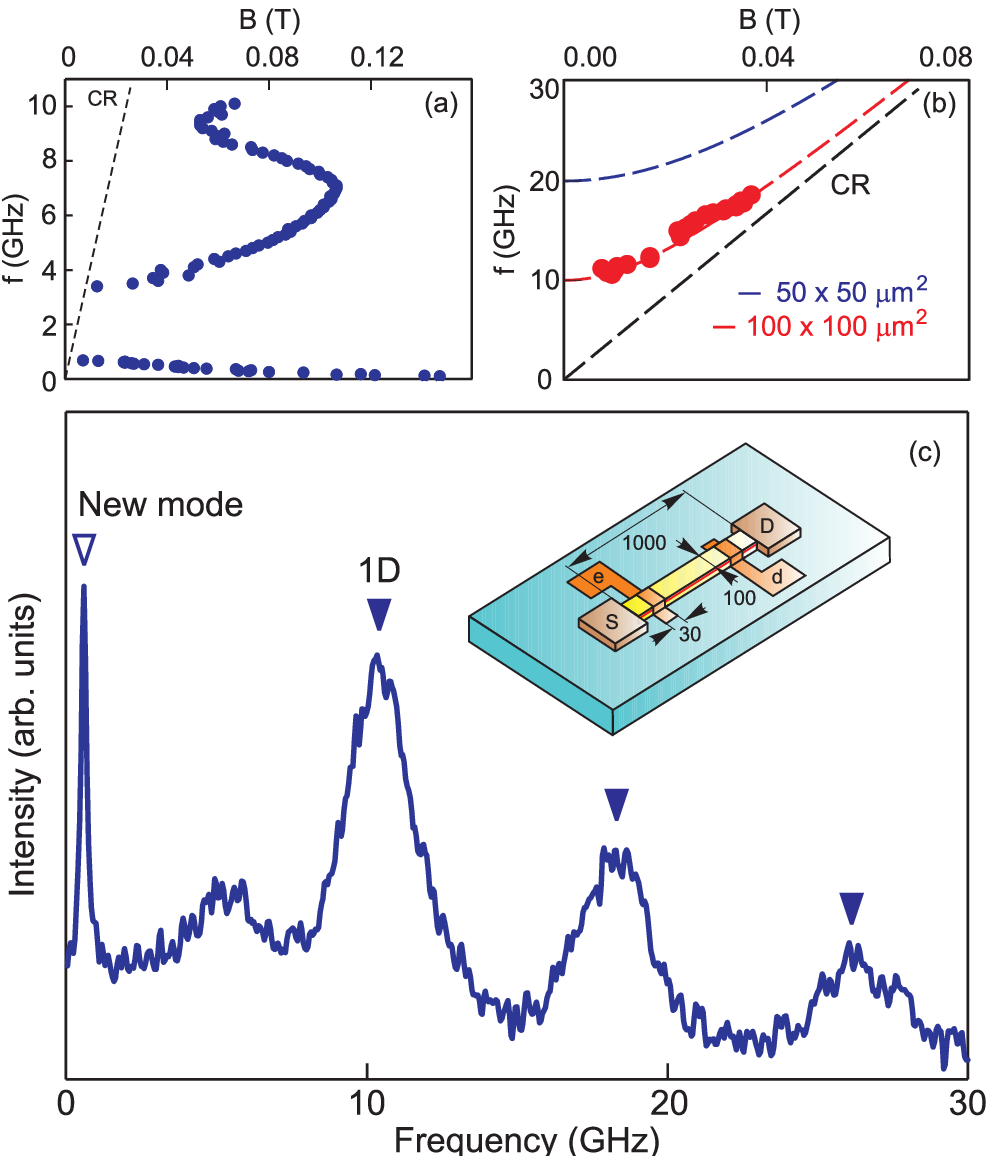}\hfill \null
\center  
\parbox[t]{0.45\textwidth}{\caption{Magnetic field dependencies of microwave transmission of sample (structure $\#7$) with $t=20$~$\mu$m, $W=50$~$\mu$m and an electron density of $n_{s}=4.3\times 10^{11}$~cm$^{-2}$ ($2\pi\sigma_{2\rm{D}}/c = 5.5$) measured for different distances $L=0.7$~mm and $0.3$~mm between gates $e$ and $d$. The plasmon magnetodispersions for these two cases are shown in the inset. The schematic view of the geometry is included.} \label{1s}}\hfill
  \parbox[t]{0.47\textwidth}{\caption{(a) Two branches of the magnetic-field position of the absorption resonances versus excitation frequency measured for structure \#1 with $t=W=50$~$\mu$m. The dashed line represents the cyclotron resonance. (b) Magnetodispersion of a screened cyclotron magnetoplasma mode measured for the sample with $t=W=100$~$\mu$m (red line). The blue line is the theoretical estimate from Eq.~(\ref{CR_T}) for the sample with $t=W=50$~$\mu$m. (c) Frequency dependency of 2DES microwave absorption for structure $\#1$. The inset shows a
schematic view of the sample geometry. All dimensions are in micron. Simultaneous excitation of new plasmon mode in the gated regions and 1D plasmon along the body of the stripe is observed.      
} \label{2s}} 
\end{figure}

\section{\textrm{III}. Dimensional plasma excitations in the structure under study}

Frequencies of all known plasma excitations, measured for the basic structure geometry and parameters are situated over $10$~GHz at $B=0$~T. As an example, Fig.~2S(b) shows the dispersion of a screened cyclotron magnetoplasma (SM) mode measured for the sample fabricated from the same wafer (\#1) with $t=W=100$~$\mu$m (Fig.~2S(c)). The frequency of the mode is described by the formula:     
\begin{equation}
\omega^2=\omega_p^2+\omega_c^2,
\label{CR_T}
\end{equation}
where $\omega_c=eB/m^{\ast}c$ is the cyclotron frequency and $\omega_p$ is given by the following expression~\cite{Chaplik:72, Fetter:86}:
\begin{equation}
\omega_p^2=\frac{4 \pi n_s e^2}{m^{\ast} \varepsilon_0} \frac{q}{1 + \varepsilon \coth qh}.
\label{Plasmon}
\end{equation}
Here, $h$ is the distance between the 2DES and the top gate layer and $q$ is the wave vector of the plasmon. The permittivity of vacuum and effective permittivity of GaAs are denoted as $\varepsilon_0$ and $\varepsilon = 12{.}8$, respectively. The red dashed line in Fig.~2S(b) marks the theoretical prediction given by Eq.~(\ref{CR_T}) for top gate size $t=W=100$~$\mu$m and distance from the gate to 2DES $h=200$~nm ($q=\sqrt{2} \pi/W$). The theory precisely describes the experimental data. The blue dashed line in Fig.~2S(b) denotes the theoretical magnetodispersion curve for the structure with gate size $t=W=50$~$\mu$m. The curve starts at $B=0$~T from a frequency of $20$~GHz, well above that of the novel observed excitation (Fig.~2S(a)).

To further illustrate our words, Fig.~2S(c) shows microwave absorption of the sample $\#1$ with $t=30$~$\mu$m, $W=100$~$\mu$m and 2DES stripe total length of $l=1$~mm. The absorption was measured using optical detection scheme. The technique is based on a comparison between luminescence spectra in the absence and presence of microwave radiation~\cite{Ashk:99}. The absorption spectrum shows a very narrow peak at $f=0{.}6$~GHz and equidistantly spaced resonances at frequencies above $10$~GHz. Narrow peak corresponds to the excitation of a novel polaritonic plasmon mode in the gated region of the 2DES. High-frequency resonances correspond to the excitation of dimensional plasma resonances~\cite{Vasiliadou, Kukushkin} with wave vector $K=N \pi/l$. Here, wave number $N$ equals $1, 2, 3 \ldots$. The dispersion of the resonances indicates longitudinal one-dimensional (1D) nature of the plasma excitation~\cite{Pinczuk, Heitmann, Kukushkin}. 

To conclude, in the regime of $\omega \tau >1$ and $2\pi\sigma_{2\rm{D}}/c>1$ novel plasmon mode and standard dimensional plasma excitations could be observed simultaneously. However, these two plasmon modes are spaced in the frequency scale by more than one order of magnitude. Therefore, for all experiments where plasmonic response in gated 2DESs was studied~\cite{Muravev, Popov} the polaritonic mode is hidden far below the available frequency diapason of the used technique.    

\section{\textrm{IV}. Density dependence of the relativistic plasmon frequency}

In Fig.~3S(a) we present magnetodispersion curves measured on three structures with different electron concentration. All experiments were conducted on samples with identical geometrical parameters for the plasmonic resonator, $W=50$~$\mu$m and $t=40$~$\mu$m. For all the structures $2\pi\sigma_{2\rm{D}}/c>1$. For each concentration, extrapolation based on Eq.~2 of magnetodispersion points is used to obtain the zero-field plasmon frequency $f_0=\omega_0/2 \pi$. The dependence of the relativistic plasmon frequency on the square root of the carrier concentration is depicted in Fig.~3S(b). Presenting the data in this manner shows that the mode frequency obeys that expected for all plasma waves in 2DESs, with a square root relation for $f_0 \sim\sqrt{n_{s}}$. Hence, the observed plasmon mode is intimately bound to the 2DES, despite its strongly delocalized polaritonic nature. 

\begin{figure}[!h]
\includegraphics[width=0.5 \textwidth]{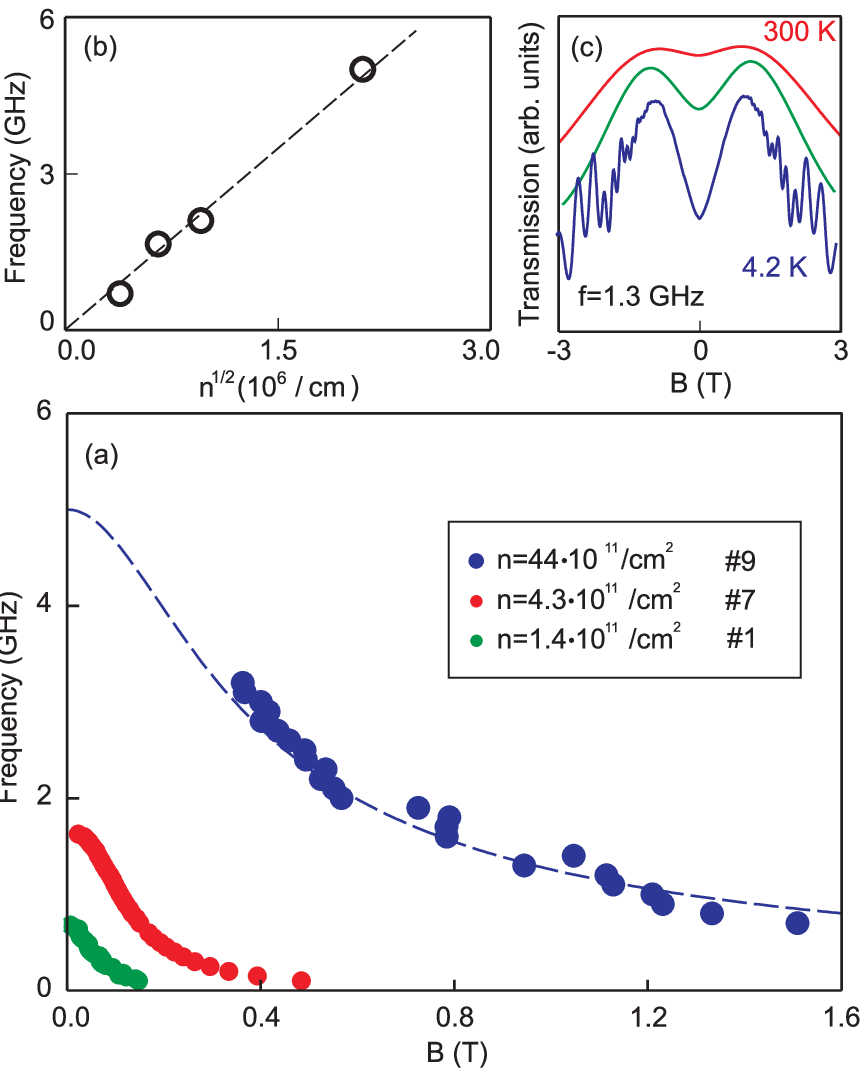}
\caption{(a) Magnetodispersion curves measured on three structures with different electron concentration. Geometrical parameters are the same for all structures $t=40$~$\mu$m, $W=50$~$\mu$m. The dashed line represents the extrapolation based on Eq.~2. (b) The zero-field limit of the plasma mode frequency depends on the square root of the electron density.
(c) Magnetic field dependencies of the microwave transmission of a sample with a record electron density of $44 \times 10^{11}/{\rm cm}^2$ (structure $\#9$) measured for three temperatures $T=4.2$~K, $160$~K, and $300$~K at a frequency of $1.3$~GHz.} 
\label{3}
\end{figure}